\documentclass[sn-mathphys,Numbered]{sn-jnl}

\usepackage{graphicx}%
\usepackage{multirow}%
\usepackage{amsmath,amssymb,amsfonts}%
\usepackage{amsthm}%
\usepackage{mathrsfs}%
\usepackage[title]{appendix}%
\usepackage{xcolor}%
\usepackage{textcomp}%
\usepackage{manyfoot}%
\usepackage{booktabs}%
\usepackage{algorithm}%
\usepackage{algorithmicx}%
\usepackage{algpseudocode}%
\usepackage{listings}%
\usepackage{bm}
\usepackage{subcaption}
\usepackage{mathabx}
\usepackage{lmodern}

\begin{document}

\title[Article Title]{Improved direction of arrival estimations with a wearable microphone array for dynamic environments by reliability weighting} 


\author[1]{\fnm{Daniel A.} \sur{Mitchell}}\email{mitchdan@post.bgu.ac.il}
\author[1]{\fnm{Boaz} \sur{Rafaely}}\email{br@bgu.ac.il}
\author[2]{\fnm{Anurag} \sur{Kumar} }\email{anuragkr90@meta.com}
\author[2]{\fnm{Vladimir} \sur{Tourbabin}}\email{vtourbabin@meta.com}

\affil[1]{\orgdiv{School of Electrical and Computer Engineering}, \orgname{Ben-Gurion University of the Negev}, \country{Israel}}

\affil[2]{\orgdiv{Reality Labs Research} \orgname{@ Meta, Menlo Park}, \country{USA}}


\abstract{
Direction-of-arrival estimation of multiple speakers in a room is an important task for a wide range of applications. In particular, challenging environments with moving speakers, reverberation and noise, lead to significant performance degradation for current methods. With the aim of better understanding factors affecting performance and improving current methods, in this paper multi-speaker direction-of-arrival (DOA) estimation is investigated using a modified version of the local space domain distance (LSDD) algorithm in a noisy, dynamic and reverberant environment employing a wearable microphone array. This study utilizes the recently published EasyCom speech dataset, recorded using a wearable microphone array mounted on eyeglasses. While the original LSDD algorithm demonstrates strong performance in static environments, its efficacy significantly diminishes in the dynamic settings of the EasyCom dataset. Several enhancements to the LSDD algorithm are developed following a comprehensive performance and system analysis, which enable improved DOA estimation under these challenging conditions. These improvements include incorporating a weighted reliability approach and introducing a new quality measure that reliably identifies the more accurate DOA estimates, thereby enhancing both the robustness and accuracy of the algorithm in challenging environments.
}





\maketitle

\section{Introduction}\label{sec1}

Localizing multiple sound sources recorded with a microphone array in a reverberant environment is an important task in a wide range of applications, including robot audition and video conferencing.  There is a wide range of direction-of-arrival (DOA) estimation algorithms available in the literature. These include beamforming \cite{van1988beamforming}, subspace methods such as multiple signal classification (MUSIC) \cite{schmidt1986multiple}, and algorithms based on time-delay of arrival estimation \cite{knapp1976generalized}. DOA estimation methods specifically developed for speech signals exploit the non-stationarity and sparsity of speech in the short-time Fourier transform (STFT) domain and enable DOA estimation even for under-determined systems with more sources than microphones \cite{reju2018efficient,brendel2018stft}. However, for noisy and reverberant environments, room reflections and additive noise mask the direct sound that  carries DOA information, thus degrading the DOA estimation performance.
\par 
Recently, a general method for DOA estimation of multiple speakers that is robust against reverberation, has been developed \cite{nadiri2014localization,rafaely2018speaker}. This method processes the signals in the time-frequency domain  
and employs a direct-path dominance (DPD) test \cite{hafezi2017multiple,tourbabin2014speaker,nadiri2014localization,rafaely2018speaker}  to identify time-frequency bins which are dominated by the direct path sound.  This  method has  been widely studied \cite{nadiri2014localization,moore2015direction,pavlidi20153d,delikaris20163d, madmoni2018improved,rafaely2018speaker,beit2019binaural} assuming a static environment, where both the sound sources and the microphone array are stationary. Among the algorithms using this method is the local space domain distance (LSDD) algorithm \cite{tourbabin2018space,mitchell2023study,tourbabin2014speaker}. This is a computationally efficient algorithm which has proven successful for multi-speaker DOA estimation in static, noisy and  reverberant environments . 
However, the DPD time-frequency algorithms, and specifically, the LSDD algorithm,   have been less intensively studied in  more realistic, dynamic environments where speakers  and/or the microphone array are moving.
\par 
In dynamic environments, the motion of the sound sources and/or microphone array may lead to a rapid change of DOAs in time. Thus, accurately tracing the DOA of speakers requires short intervals between successive DOA estimates. Additionally, the DOA estimates may be smoothed in time using a tracking algorithm. Although DOA estimation and tracking algorithms in dynamic environments have been the subject of several recent studies \cite{slam2018,wang2013high,rusrus2023moving,hammer2021dynamically}, including studies presented as part of the Acoustic Source Localization and Tracking (LOCATA) \cite{lollmann2018locata,lebarbenchon2018evaluation,madmoni2018description} challenge, none of these have included experiments with wearable microphone arrays \cite{zhao2023audio}. 
Such scenes are becoming increasingly more popular due to the increased interest in applications involving augmented reality \cite{carmigniani2011augmented}. 
\par
In this article we address the problem of DOA estimation in a realistic scenario involving a wearable microphone array. The algorithm chosen for this work was based on the LSDD algorithm. As part of this effort we developed a framework for evaluating DOA estimates which  change over time and for optimizing the many inter-dependent parameters in the DOA estimation algorithm. DOA estimates are obtained in short intervals for each active speaker to account for the rapid change in DOA. The DOA estimates were calculated using a subtractive weighted cluster algorithm, whose outputs include both the final DOA estimated within a time interval, and their corresponding estimation quality.  Finally, we propose a reliability-weighted LSDD algorithm for improved robustness under these challenging conditions.
 In summary, the contributions of this article are:

    \textit{Development of a Novel Methodology for Increasing Robustness under Noisy and Dynamic Environments}.   
    The increased robustness is achieved by incorporating an array reliability measure and a cluster quality measure. The latter is integrated into a  clustering algorithm that enhances the accuracy and robustness of the final DOA estimates within a time interval. The weighted reliability measure emerges from array characteristics and is also integrated into the DOA estimation algorithm. Both measures are detailed next. 

\textit{Array  Reliability measure}. A mathematical model for analyzing the response of the microphone array steering vectors is described. generating a universal directivity map (UDM). The UDM provides an array-based reliability weight for each time-frequency DOA estimate. This reliability weight reflects the spatial selectivity of the array as a function of direction and frequency.

    \textit{Cluster  Quality measure}. A novel quality measure is calculated for each final DOA estimate within a time interval.  The quality measure is based on the relative weights of individual clusters, and is shown to be highly correlated with the DOA accuracy, therefore supporting an improved selection of a subset of accurate DOA estimates.
\par

The experiments described in the article were performed using the Easy Communication (EasyCom)  \cite{donley2021easycom} dataset which was explicitly designed to represent a realistic cocktail-party environment.  The proposed DOA estimation algorithm was found to have an improved performance on the EasyCom dataset while retaining the intrinsic speed and simplicity of the baseline LSDD algorithm.   
\section{Signal Model} 
 In this section we discuss the mathematical model underlying the
 LSDD algorithm \cite{tourbabin2018space} for  DOA  estimation.
Assume a microphone array comprising   $M$ microphones arranged according to some pre-defined configuration. Next, consider a sound field comprised of $K$ plane waves generated by the same number of far-field 
sources, arriving from directions $\Psi_{k}, k \in \{1,2,\ldots,K\}$. These sources represent the direct sound from speakers in the scene, as well as reflections due to objects and room boundaries. 
\par
The signals captured by the microphones in the array transformed into the joint time-frequency domain by applying the short-time Fourier transform (STFT). This is done by first separating the speech signal into short time intervals of length $\delta t$. A fast Fourier transform (FFT) is then applied to each time segment. Following these pre-processing steps, the signal received by the $m$th microphone in time-frequency bin $(t,f)$ can be described as
\begin{equation}
{x}_{m}(t,f) =  \sum_{k=1}^{K}s_{k}(t,f)v_{m}(\Psi_{k}(t),f)+
n_{m}(t,f)\;,
\label{eq:equation1}
\end{equation}
where $v_{m}(\Psi_{k}(t),f)$ denotes the response of the $m$th microphone  to a unit-amplitude plane wave at frequency $f$ arriving from the $k$th source in direction $\Psi_{k}(t)$; 
$s_{k}(t,f)$ is  the amplitude of the $k$th sound source signal; and
$n_{m}(t,f)$ is the additive noise present at the $m$th microphone. In writing Eq. (\ref{eq:equation1}) we have assumed  the  multiplicative transfer function property holds \cite{MTF} and that $\Psi_{k}(t)$ is effectively constant within the duration of $\delta t$. 
\par 
The corresponding complex signal received by the entire microphone array is thus
\begin{align}
\textbf{x}(t,f) &= [x_{1}(t,f), x_{2}(t,f),\ldots,x_{M}(t,f)]^{T}\;,\nonumber\\   &=\sum_{k=1}^{K}s_k(t,f)\textbf{v}(\Psi_{k}(t),f)+
\textbf{n}(t,f)\;,
\label{eq:equation1a}
\end{align}
where
\begin{equation}
\begin{split}
\textbf{v}(\Psi_{k}(t),f)&=[v_1(\Psi_{k}(t),f),v_2(\Psi_{k}(t),f),\ldots,v_M(\Psi_{k}(t),f)]^{T} \;,\\ 
\textbf{n}(t,f)&=[n_1(t,f), n_2(t,f),\ldots,n_M(t,f)]^{T}\;.
\end{split}
\end{equation}

\section {DOA Estimation Baseline }\label{sect2B}
In this section, the \emph{baseline} 
LSDD algorithm for DOA estimation is described. 
 This algorithm \cite{anand2016comparison, rafaely2018speaker, mitchell2023study}
has been recently developed, demonstrating computational simplicity and relative robustness to reverberation.
\par 
The essential element of the algorithm
 \cite{tourbabin2018space} is the directional spectrum $\textbf{S}(t,f)$, which is defined for each time-frequency bin $(t,f)$ over a grid of $L$ directions $\phi_{l},l \in \{1,2,\ldots,L\}$:
 \begin{align}
\textbf{S}(t,f)&=[S_{1}(t,f) S_{2}(t,f),\ldots,S_{L}(t,f)]^{T}\;. \label{eqn_entropyS}
\end{align}
The components $S_{l}(t,f)$ are calculated by measuring the similarity between the vector $\textbf{x}(t,f)$ and  the vector $\textbf{v}(\phi_{l},f)$ (in the direction $\phi_{l}$).  Thus
\begin{align}
    S_{l}(t,f)&=d \left( \textbf{x}\left(t,f\right), \textbf{v}\left(\phi_{l},f\right) \right) \;
\label{eq:equation2}
\end{align}
where $\textbf{v}(\phi_{l},f) = [v_1(\phi_{l},f),\ldots,v_M(\phi_{l},f)]^{T}$ and
 $d(\textbf{a},\textbf{b})$ denotes 
the similarity between two vectors $\textbf{a}$ and $\textbf{b}$.
\par 
Originally, in the LSDD algorithm 
\cite{tourbabin2018space}, 
$d(\textbf{a},\textbf{b})$ was defined as
\begin{equation}    d(\textbf{a},\textbf{b})=\frac{1}
{{\underset{\beta}\min}\big 
(\frac{\|\textbf{a}-\beta\textbf{b}\|}
{\|\textbf{a}\|}\big)}\;.
    \label{eq:equation3}
\end{equation}

However, following \cite{mitchell2023study}  
the more conventional cosine similarity measure is used:
\begin{equation}
d(\textbf{a},\textbf{b}) = \frac{|<\textbf{a},\textbf{b}>|}{\|\textbf{a}\|\|\textbf{b}\|}\;,
\label{eq:equation4}
\end{equation}
where $<\textbf{a},\textbf{b}>$ denotes the inner product between $\textbf{a}$ and $\textbf{b}$.

\par Given the 
spectrum  
$\textbf{S}(t,f)$, 
the estimated DOA  for the time-frequency bin $(t,f)$ is defined as 
\begin{equation}\label{eqn_est_phi_tfbin}
    \hat{\phi}(t,f)=\phi_{l_{\rm max}}\;,
    \end{equation}
    where 
    \begin{equation}\label{eqn_est_phi_tfbin_optimum}
    l_{\rm max}={\arg \underset{l}\max}\big(S_{l}(t,f)\big)\;.
\end{equation}
By definition, the angle $\hat{\phi}(t,f)$ is measured \emph{relative to the microphone array axis}.
In this work, because arrays may move,  the DOA is measured relative to an axis fixed to the room. In this case, the corresponding DOA estimate is denoted as $\hat{\theta}(t,f)$, where
\begin{equation}\label{eqn_theta_tf_from_phi_tf}
    \hat{\theta}(t,f)=\hat{\phi}(t,f)+\delta_{\text{array}}\;,
\end{equation}
and $\delta_{\text{array}}$ is the angle between the microphone array axis and the axis fixed relative to the room. 
\par

Some of the bins $(t,f)$ do not, however, contain a valid 
$\hat{\theta}(t,f)$ estimate.  These are bins in which the direct signal from the speaker is masked by noise and/or reverberations. We may eliminate these bins by calculating a direct-path dominance (DPD) measure $\xi(t,f)$ for each bin $(t,f)$ which we then test against a threshold $\lambda$. We use the binary function to identify \lq\lq valid" bins:
\begin{align} \label{eq:identity}   
V(t,f)&= 
\begin{cases}
    1\;\; (\text{\lq\lq valid"})& \text{if } \xi(t,f)\ge \lambda\;,\\
    0              & \text{otherwise}\;,
\end{cases}
\end{align}
\par 
There are various DPD measures available for identifying direct-sound dominant time-frequency bins \cite{hafezi2017multiple,tourbabin2014speaker,nadiri2014localization,rafaely2018speaker}. 
The chosen DPD formula is as per \cite{tourbabin2018space}:
\begin{equation}
\xi(t,f) = \underset{l}{ \max}\big(S_l(t,f)\big)\;.
\label{eq:equation51}
\end{equation}

\par
In practice, following \cite{beit2020importance}, the spectrum $\textbf{S}(t,f)$ is smoothed before calculating $\hat{\theta}(t,f)$ and $\xi(t,f)$. The  
smoothed spectrum $\overline{\textbf{S}}(t,f)$ is obtained by 
averaging $\textbf{S}(t,f)$ in the range $t\in R_{t}$ and $f \in R_{f}$ around $(t,f)$. 
\par
Consideration is now given to how individual time-frequency estimates $\hat{\theta}(t,f)$ can be combined to provide accurate DOA estimates, first in a static environment and then in a dynamic environment.

\subsection{DOA Estimation in a Static Environment}
The problem of using the baseline LSDD algorithm for DOA estimation in a static environment is now considered. In such an environment, both the sound sources and the microphone array remain stationary. Under these conditions, accurate DOA estimates can be achieved by averaging the valid time-frequency DOA estimates $\hat{\theta}(t,f)$  over an extended period \cite{tourbabin2018space,clust-araki2011doa}. 
 \par 
 Let us assume there are $K$ speakers. The individual estimates $\hat{\theta}(t,f)$ are then clustered into $K$ clusters \cite{yager1994clustering}. Assuming a different cluster for each speaker, we may generate a DOA estimate $\hat{\Theta}(k),k\in\{1,2,\ldots,K\}$, for each speaker $k$ by averaging the valid estimates $\hat{\theta}(t,f)$ for each cluster: 
\begin{equation}
    \widehat{\Theta}(k) = \frac{1}{W_{k}} \sum_{(t,f)} \hat{\theta}(t,f) V(t,f) C_{k}(t,f)\;,
\label{eqn_18zz}
\end{equation}
where
\begin{align}\label{Eqn_18zz_supplement}
    W_{k}&=\sum_{(t,f)}V(t,f)C_{k}(t,f)\;,\\
C_{k}(t,f)&= 
\label{Eqn_cluster_indicator_function}
\begin{cases}
    1 & \text{if } (t,f)\;\; \text{belongs to the $k$th cluster}\;,\\
    0              & \text{otherwise}\;,
\end{cases}
\end{align}
\subsection{DOA Estimation in a Dynamic Environment}\label{Dynamic_DOA_forBaselie}
In this section, the problem of performing DOA estimation in a dynamic environment is considered. In such environments, where the positions of sound sources and/or the microphone array change over time, traditional long-term averaging of individual time-frequency DOA estimates is not feasible. This challenge arises from the necessity to continuously track the changing directions of sound sources. To accommodate these challenging conditions, the approach used in static scenarios requires modification.
\par
The primary modification involves segmenting the time axis into short intervals, $\Delta T$, during which the sound sources and microphone array can be considered approximately spatially constant. Each interval is labeled by its mid-time $T$, and the aim is to estimate a final DOA $\hat{\Theta}_{T}(k)$ for each active speaker $k$, defined as a speaker who is vocal during the interval $T$. 
\par
The application of the baseline LSDD algorithm for DOA estimation in a dynamic environment is now described. Fig. \ref{fig:flowchart-static}  shows the corresponding flow chart in the form of a  block diagram.  The main blocks in Fig. \ref{fig:flowchart-static} are:
 \begin{figure}[H]
    \centering
    \includegraphics[width=0.9\linewidth]{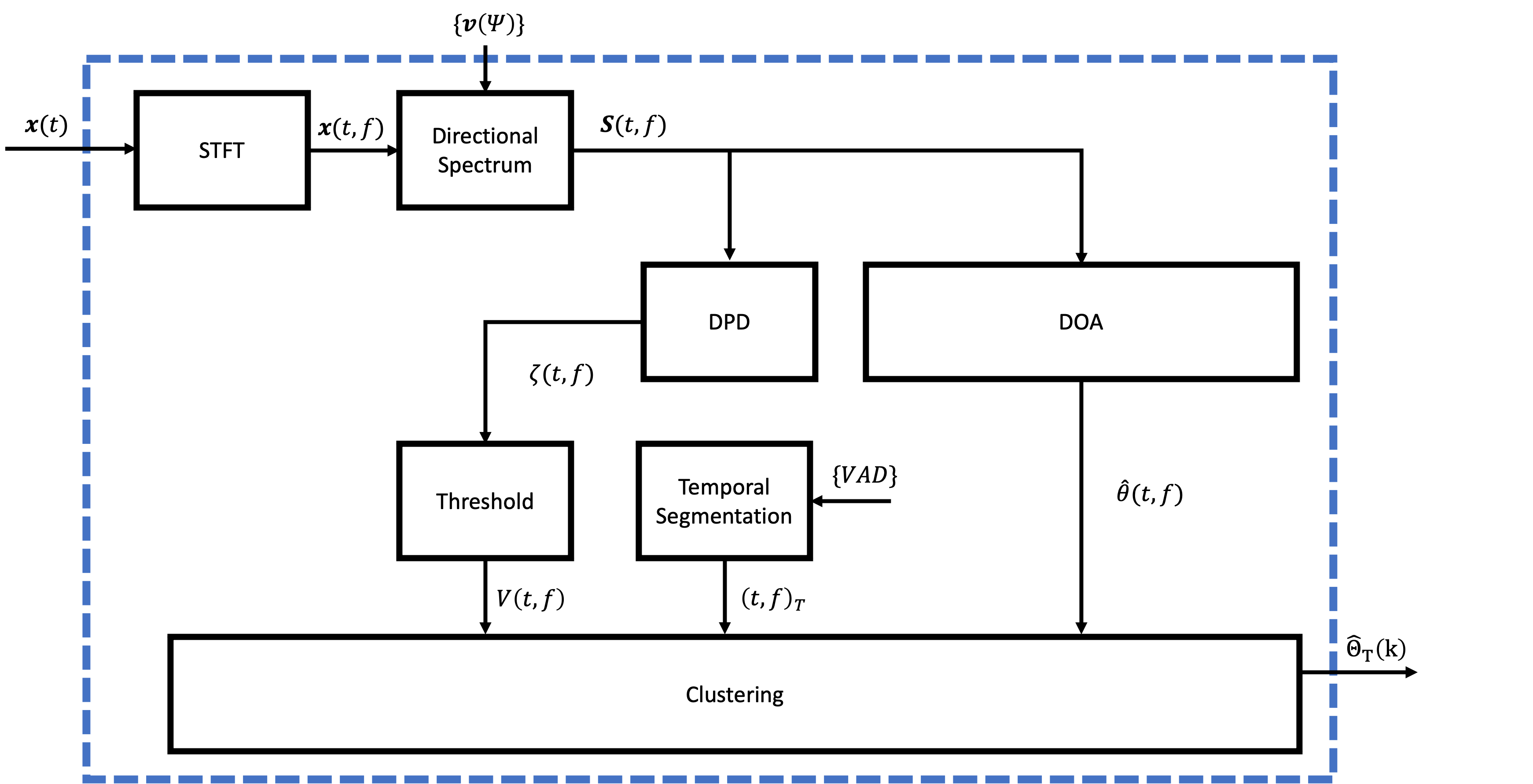}
    \caption{Flowchart showing the main processing blocks for DOA estimation using LSDD-base algorithm in a dynamic environment. 
    }
    \label{fig:flowchart-static}
\end{figure}
\begin{itemize}
   
     \item \textbf{STFT}.
        The microphone signals are transformed into the time-frequency domain using the short-time Fourier transform. The size of each time-frequency bin is denoted as $(\delta t,\delta f)$.  Initially, a characteristic frequency range for the microphone array is defined as between $f_{\text{low}}$ and $f_{\text{high}}$. Bins at frequencies outside this range are excluded from further processing. More details regarding the selection of $f_{\text{low}}$ and $f_{\text{high}}$ will be discussed in section \ref{SECT_freq_range}. 
     \item \textbf{Directional Spectrum} $\textbf{S}$. The spectrum $\textbf{S}(t,f)$ is calculated around each time-frequency bin $(t,f)$ using Eq. (\ref{eq:equation2}). A smoothed spectrum $\overline{\textbf{S}}(t,f)$ is calculated by averaging $\textbf{S}(t,f)$ in the range $t \in R_t$ and $f \in R_f$ around $(t,f)$. Further details regarding the averaging will be discussed in section \ref{sec:freqSmooth}. The importance of smoothing the spectrum is shown in Ref. \cite{beit2020importance}. 
     \item
     \textbf{DOA Estimate} $\hat{\theta}(t,f)$. For each bin $(t,f)$, the DOA estimate is calculated using the smoothed spectrum $\overline{\textbf{S}}(t,f)$ as in Eqs.  (\ref{eqn_est_phi_tfbin}) and (\ref{eqn_est_phi_tfbin_optimum}). 
    \item
    \textbf{DPD Measure} $\xi(t,f)$. For each bin $(t,f)$, the DPD measure $\xi(t,f)$ is calculated using the smoothed spectrum $\overline{\textbf{S}}(t,f)$ in Eq. (\ref{eq:equation51}). 
    \item
    \textbf{Threshold} $\lambda$. All bins $(t,f)$ with DPD value $\xi(t,f)$  that exceed the threshold $\lambda$ are identified. These bins are denoted as \lq\lq valid" and are labeled $V(t,f)=1$ (Eq. (\ref{eq:identity})).
      \item \textbf{Active Speech Time Segmentation}. The time axis is divided into intervals of length $\Delta T$. If $T$ denotes the mid-time of a given interval, then the interval $T$ is classified as active if one or more speakers are speaking throughout the entire interval. In practice, the activity of $T$ is provided by information from a voice activity detector (VAD).
    \item  \textbf{Clustering }.  All valid bins in the interval $T$ are clustered. For each cluster $k$,  the average DOA estimate $\widehat{\Theta}_{T}(k), k\in \{1,2,\ldots,K\}$, is found (Eq. (\ref{eqn_18zz})). 
\end{itemize} 
\par

   \section{Proposed DOA Estimation Algorithm}\label{SECT_newLSDD_algorithms}
   In this section, the proposed DOA estimation algorithm is described, incorporating two improvments to the baseline algorithm. First, a DOA reliability weight $w(t,f)$ for each time-frequency bin $(t,f)$ is introduced. This allows DOA estimates that are more reliable to contribute more to the final DOA estimates. The reliability-weighted DOA estimate for the $k$th speaker is mathematically given by:
   \begin{equation}
    \widehat{\Theta}(k) = \frac{1}{W_{k}} \sum_{(t,f)} w(t,f)\hat{\theta}(t,f) V(t,f) C_{k}(t,f)\;,
\label{eqn_18zz_new}
\end{equation}
where
\begin{equation}\label{Eqn_18zz_supplement_new}
W_{k}=\sum_{(t,f)}w(t,f)V(t,f)C_{k}(t,f)\;,
\end{equation}
 The  reliability weight $w$ is discussed in  Sect. \ref{SECT_new_reliabity_weight}.
 
 Second, a quality measure $Q(k)$ for each cluster $k\in \{1,2,\ldots,K\}$ is introduced, assessing the likelihood that a cluster is a \lq\lq true" cluster rather than a spurious one due to noise and interference. As will be demonstrated in Sect. \ref{SECT_results}, the quality measure is directly related to the expected accuracy of the final DOA estimates $\hat{\Theta}(k)$. Using $Q(k)$, DOA estimates with high accuracy can be selected, and less accurate estimates can be discarded.  Alternatively, $Q(k)$ may be employed as a weight in a  subsequent DOA tracking algorithm.  The quality $Q$ is  discussed in Sect. \ref{SECT_NewQ}.
 \par
We denote the new DOA algorithm as LSDD-wQ to emphasize that it incorporates a reliability weight $w(t,f)$ and a cluster quality $Q$. 
 
\subsection{\texorpdfstring{DOA Reliability Weight $w$}{DOA Reliability Weight w}} \label{SECT_new_reliabity_weight} 

The reliability weight $w(t,f)$ is modeled  
as a product of two reliability factors or functions:
\begin{itemize}
    \item Data-based reliability $\xi(t,f)$. This function relies on the strength of the DPD measure $\xi(t,f)$  and represents the likelihood that the $(t,f)$ bin is dominated by the direct-path. 
    \item Array-based reliability $\alpha$. This function measures the microphone directivity at any given look direction and frequency, calculated \emph{directly} from the microphone array characteristics. The array-based reliability function   does \emph{not} change with the scenario and  needs only to be computed once for each microphone array.
    \end{itemize}

   In calculating $\alpha$ it is supposed that it is a function of the estimated DOA $\hat{\phi}(t,f)$ and the frequency $f$. Thus, the reliabilty weight $w(t,f)$ can be expressed as:
   \begin{equation}
   w(t,f)=\alpha(\hat{\phi}(t,f),f)\times \xi(t,f) \;.
   \end{equation}
\par
Consider a specific steering vector $\textbf{v}(\phi_{h},f)$, corresponding to a frequency $f$ and direction $\phi_{h}$. The microphone array directivity for $\phi_{h}$ is then related to the similarity between $\textbf{v}(\phi_{h},f)$ and the set of steering vectors $\textbf{v}(\phi_{l},f), l \in \{1,2,...,L\}$. This calculation is repeated for all frequencies $f$ yielding the following measure:
\begin{equation}\label{EQN_STEERING+VECT_SIM}
    {\Lambda}_{h}(\phi_l,f) = d(\textbf{v}(\phi_{h},f),\textbf{v}(\phi_{l},f))\;,
\end{equation}
where Eq. (\ref{eq:equation4}) is used to compute 
$d(\textbf{v}(\phi_{h},f),\textbf{v}(\phi_{l},f))$. Physically $\Lambda_{h}(\phi_{l},f)$ 
is related to the directivity of the microphone array  at an angle $\phi_{h}$. 
\par
In \emph{deriving} the array-based reliability $\alpha$  we require $\alpha(\phi_{h})$ to have a high value if $\Lambda_{h}(\phi_{l})$ is high when  $\phi_{l}\sim \phi_{h}$ and $\Lambda_{h}(\phi_{l})$ is low  elsewhere.  We may implement this requirement  using the directivity factor \cite{beranek2012acoustics}.  However, in practice, we found using   the following mathematical model to be more robust.
\par 
First $\Lambda_h( \phi_l,f)$ is binarized using a predefined threshold $thr$:
\begin{equation}\label{EQN_binarizationOfLambda}
    B_h(\phi_l,f) = \begin{cases}
        1 & \text{if } \Lambda_h( \phi_l,f) > thr\;,\\
        0 & \text{otherwise}\;.
    \end{cases}
\end{equation}

Then, for each frequency $f$ we count 
\begin{itemize}
    \item 
$m_h(f)$. This is the number of   bins that are near $\phi_h$. Physically $m_h(f)$ represents the  concentration of $\Lambda_{h}$ around the angle $\phi_{h}$ at the frequency $f$. 
\item $M_h(f)$. This is the number of bins that are far from $\phi_{h}$. Physically $M_h(f)$  represents the number of outliers and sidelobes in $\Lambda_{h}$  which are far from $\phi_{h}$ at the frequency $f$. 
\end{itemize}
Mathematically, $m_h(f)$ and $M_h(f)$ are defined as follows:
\begin{eqnarray}
    m_h(f) &=& \sum_{|\phi_l - \phi_h| < \delta_{\text{near}}} B_h( \phi_l,f)\;,\\
    M_h(f) &=& \sum_{|\phi_l - \phi_h| > \delta_{\text{far}}} B_h( \phi_l,f)
\end{eqnarray}
where  $\delta_{\text{near}}$ and $\delta_{\text{far}}$  are, respectively, a small angle ($\approx 10$ deg) and  a large angle ($\approx 25$ deg).
\par 
 We now combine $m_h(f)$ and $M_h(f)$, in such a way that we emphasize  frequencies which  \emph{simultaneously} have high $m_h(f)$ values and low $M_h(f)$ values. A  method of doing this is to   sort  $m_h(f)$ over all frequencies $f$  in ascending value and sort $M_h(f)$ over all frequencies $f$ in  descending value. We  then add the corresponding indices, $r_{h}(f)$ and $R_{h}(f)$, in the sorted arrays:
\begin{equation}
    \mathcal{R}_{h}(f)=r_{h}(f)+R_{h}(f)\;.
    \end{equation}
    where $r_{h}(f)=l$ if $m_{h}(f)$ is the $l$th largest $m_{h}$ value and
    $R_{h}(f)=l$ if $M_{h}(f)$ is the $l$th smallest $M_{h}$ value.
    \par
 We now
normalize the $\mathcal{R}_{h}(f)$ values into a  \emph{universal directivity map} (UDM) $\Xi$.
     Let $\mathcal{R}_{\text{max}}$ and $\mathcal{R}_{\text{min}}$, denote the corresponding maximum and minimum $\mathcal{R}$ values. 
    Then $\Xi(\phi_{h},f)$ is defined as the normalized rank 
    \begin{equation}
        \Xi(\phi_{h},f)=\frac{\mathcal{R}_{h}(f)-\mathcal{R}_{\text{min}}}
        {\mathcal{R}_{\text{max}}-\mathcal{R}_{\text{min}}}\;.
        \end{equation}
The UDM is used directly as the array-based reliability function $\alpha$. Thus:

\begin{equation}
    \alpha(\hat{\phi}(t,f),f)=\alpha(\phi_{h},f)\;,
    \end{equation}
where $(\phi_{h},f)$ is the point closest to $(\hat{\phi},f)$.

\subsection{\texorpdfstring{Cluster Quality $Q$}{Cluster Quality Q}} \label{SECT_NewQ}

The quality measure $Q$ measures the likelihood that a cluster of a set of DOA estimates $\hat{\theta}(t,f)$  is a true cluster and not a spurious cluster due to noise or interference. Although the formula developed for $Q$ is a general formula it takes on a particularly simple form when used with a subtractive clustering algorithm. In what follows we shall therefore assume the DOA estimates $\hat{\theta}(t,f)$ are clustered using a subtractive clustering algorithm \cite{yager1994clustering}.
\par
The subtractive weighted cluster algorithm works \emph{iteratively} such that one cluster is generated per iteration. The clusters are ordered according to their weight $W_{k}$ (Eq. (\ref{Eqn_18zz_supplement_new})), where $W_{k}> W_{l}$ if $l>k$. Assuming $K$ active speakers are present  (this information is provided by a Voice Activity Detector), the cluster algorithm runs for $K$ iterations and the corresponding DOA estimates $\hat{\Theta}(k), k\in\{1,2,\ldots,K\}$, are calculated using Eq. (\ref{eqn_18zz_new}).
\par
Each estimate $\widehat{\Theta}(k)$ is assigned a  quality measure $Q(k)$ by comparing the weight $W_{k}$ with the $(K+1)$th weight $W_{K+1}$     
\begin{equation}\label{eq:quality_measure_Q}
    Q(k) = \frac{W_k}{W_{K+1}}\;.
\end{equation}
\par 
The equation for the new quality measure is adopted from Lowe \cite{lowe2004quality} who used  similar reasoning to  define  keypoint quality in image registration. 
\emph{Note}: to calculate $Q(k)$ we need to run the cluster algorithm for one more iteration, i.~e. $(K+1)$ iterations altogether. 

\par
The pseudo-code used for clustering the time-frequency estimates $\hat{\theta}(t,f)$ and estimating  $\hat{\Theta}(k)$ and the quality $Q(k)$ is detailed in Alg. 
 \ref{ALG_pseudocode_clustering}. 
\begin{algorithm}[H]
\caption{Subtractive Weighted Clustering for DOA Estimates and their Qualities}\label{ALG_pseudocode_clustering}
\begin{algorithmic}[1]
\For{each interval $T$}
    \State Initialize histogram $H(i)$ for $i \in \{-180, -179, \ldots, +180\}$
    \For{each valid time-frequency bin $(t,f)$ ($V(t,f) = 1$)}
        \State Identify the bin index \( i \) closest to \( \hat{\theta}(t,f) \) such that:
        \[
        |i - \hat{\theta}(t,f)| \leq 0.5
        \]
        \State Increment the histogram at bin \( i \) by the weight of the bin:
        \[
        H(i) \gets H(i) + w(t,f)
        \]
    \EndFor
    \For{$k = 1$ to $K+1$}
        \State Sum the values of histogram $H(i)$ within a window for smoothing:
        \[
        L(i) = \sum_{j=i-\delta}^{i+\delta} H(j)
        \]
        \State Find the cluster center $i_k$ where $L(i)$ is maximum
        \State Each valid $(t,f)$ bin which lies  within the window $(i_{k}-\delta,i_{k}+\delta)$ are assigned $\phantom{xxxxxx}$ a cluster indicator function $C_{k}(t,f)=1$ (Eq. (\ref{Eqn_cluster_indicator_function})).
        \State Calculate the sum of weights for the $k$th cluster using Eq. (\ref{Eqn_18zz_supplement_new}). 
        \State Calculate $\hat{\Theta}(k)$ 
        using Eq. (\ref{eqn_18zz_new}).
        \State Zero out histogram entries near $i_k$ to prevent them effecting  the next $\phantom{xxxxxx}$ cluster:
        \[
        H(j) = 0 \text{ for } j \in \{i_k - \delta, \ldots, i_k + \delta\}
        \]
    \EndFor
    \For{$k = 1$ to $K$} 
        \State Calculate the quality measure $Q(k)$ using  Eq. (\ref{eq:quality_measure_Q})
    \EndFor
\EndFor
\end{algorithmic}
\end{algorithm}
 We now summarize the equations underlying the new DOA estimation algorithm  which we denoted as LSDD-wQ to emphasize it is jointly characterized by a  reliability weight $w_{\text{new}}(t,f)$ and a  quality measure $Q_{\text{new}}(k)$. For each cluster $k$ the algorithm generates a DOA estimate $\widehat{\Theta}_{\text{new}}(k)$ and a quality measure $Q_{\text{new}}(k)$: 
\begin{equation}\label{family_LSDD_1}
\begin{aligned}
\text{LSDD-wQ}:\quad
\begin{cases}
    \widehat{\Theta}_{\text{new}}(k)&=\displaystyle\frac{1}{W_{k}}\sum_{(t,f)}w_{\text{new}}(t,f)\hat{\theta}(t,f)V(t,f)C_{k}(t,f)\;, \\
    Q_{\text{new}}(k)&=\displaystyle\frac{W_{k}}{W_{K+1}}\;,
    \end{cases}
    \end{aligned}
    \end{equation}
where
\begin{equation}\label{family_LSDD_2}
\begin{aligned}
w_{\text{new}}(t,f)&=\alpha(\hat{\phi},f)\times \xi(t,f)\;,\\
    W_{k}&=\sum_{(t,f)}w_{\text{new}}(t,f)V(t,f)C_{k}(t,f)\;.
\end{aligned}
\end{equation}
\par
For convenience, the baseline LSDD algorithm (Eqs. (\ref{eqn_18zz}) and (\ref{Eqn_18zz_supplement}) is rewritten using the same notation.  In the baseline algorithm, all time-frequency bins are assigned a uniform weight $w_{\text{base}}(t,f)=1$ and all the clusters are assumed to be \lq\lq true", each with a quality measure $Q_{\text{base}}=1$. Thus, for each cluster $k$, the baseline algorithm (denoted hereinafter as LSDD-base)  generates  a DOA estimate $\widehat{\Theta}_{\text{base}}(k)$ and quality measure $Q_{\text{base}}(k)$:
\begin{equation}\label{family_LSDD_1base}
\begin{aligned}
\text{LSDD-base}:\quad\begin{cases}
    \widehat{\Theta}_{\text{base}}(k)&=\displaystyle\frac{1}{W_{k}}\sum_{(t,f)}w_{\text{base}}(t,f)\hat{\theta}(t,f)V(t,f)C_{k}(t,f)\;, \\
    Q_{\text{base}}(k)&=1\;,
    \end{cases}
    \end{aligned}
    \end{equation}
where
\begin{equation}\label{family_LSDD_2base}
\begin{aligned}
w_{\text{base}}(t,f)&=1\;,\\
    W_{k}&=\sum_{(t,f)}w_{\text{base}}(t,f)V(t,f)C_{k}(t,f)\;.
\end{aligned}
\end{equation}

\subsection{Proposed Algorithm for Dynamic Scenarios}
\begin{figure}[H]
    \centering
    \includegraphics[width=0.9\linewidth]{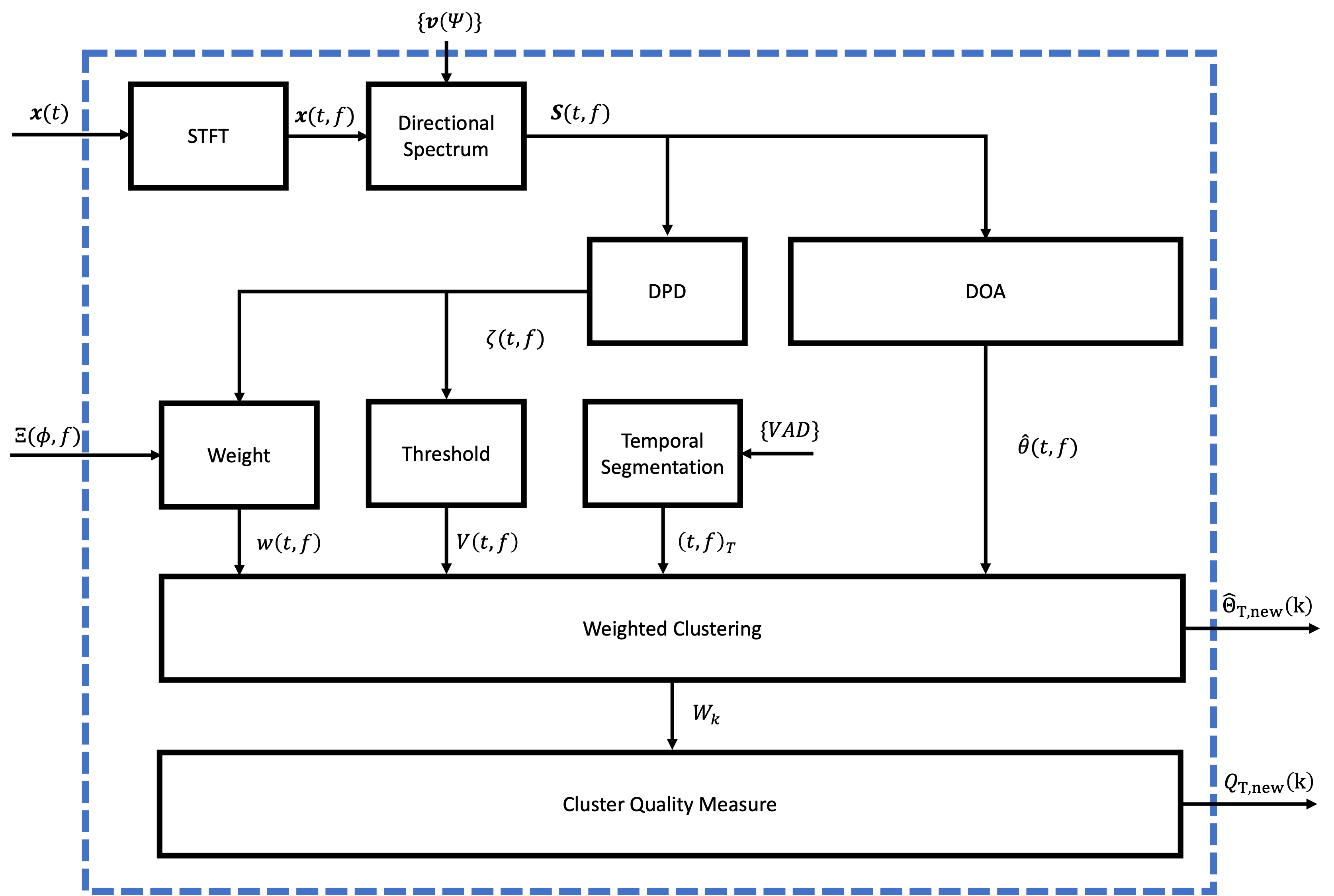}
    \caption{Flowchart showing the main processing blocks for DOA estimation calculation in a dynamic environment using the $\text{LSDD-wQ}$ algorithm.}
    \label{fig:flowchart-dynamic}
\end{figure}

In this section, the adaption of  LSDD-wQ to a dynamic environment is considered. We follow the same approach described in Sect. \ref{Dynamic_DOA_forBaselie}. The corresponding
  flowchart  is shown in Fig. \ref{fig:flowchart-dynamic}. This represents a significant modification of the baseline flowchart shown in Fig. \ref{fig:flowchart-static}.

Apart from the modules which appear in Fig. \ref{fig:flowchart-static}, the new flowchart in Fig. \ref{fig:flowchart-dynamic} includes three new signal processing modules. A list of these modules, together with a brief description of their content, is now given:
\begin{itemize}
    \item \textbf{DOA Reliability Weight} $w_{\text{new}}(t,f)$. For each time-frequency bin, calculate a weight $w_{\text{new}}(t,f)$ which represents the reliability of its DOA estimate, using Eq. (\ref{family_LSDD_2}).
    \item \textbf{Subtractive Weighted Clustering}. For each active interval $T$, all valid bins, i.e. bins that satisfy $V(t,f)=1$, are clustered by applying a \emph{subtractive weighted} clustering algorithm (see Alg. \ref{ALG_pseudocode_clustering}). 
 The output of the algorithm are a set of DOA estimates $\widehat{\Theta}_{T,\text{new}}(k)$ (Eq. \ref{family_LSDD_1}).
 \item
\textbf{Cluster Quality Measure}. For each cluster $k, k\in \{1,2,\ldots,K\}$, found in time interval $T$, calculate a quality measure  $Q_{T,\text{new}}(k)$ (Eq. \ref{family_LSDD_1}).
\end{itemize}

\section{Data and Array Analysis}
In this section, the EasyComm dataset and the microphone array on which the DOA experiments were performed are presented, along with their characteristics. This characterization will support parameter selection for the algorithms.

\subsection{The EasyComm dataset}\label{sub-easycom}
\begin{figure}[H]
    \centering
    \includegraphics[scale=0.08]{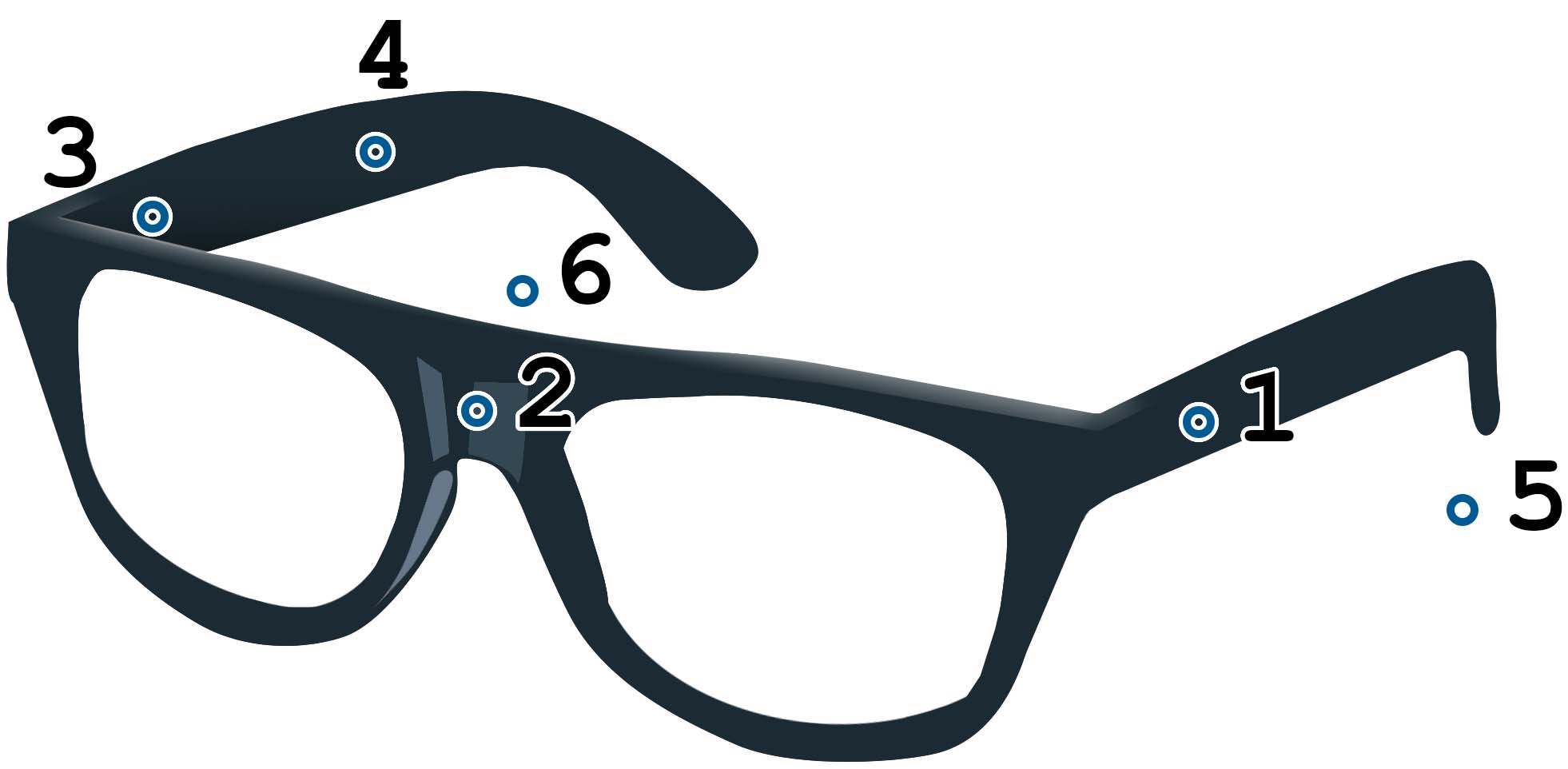}
    \caption{Illustration of the AR glasses with locations of microphones  \cite{donley2021easycom}.  Four of the microphones are fixed rigidly to the glasses and two of the microphones are placed in the user's ears.}
    \label{fig:glasses}
\end{figure}

The EasyComm dataset \cite{donley2021easycom} incorporates a scenario with several speakers having a conversation, with the audio signals captured by augmented reality (AR) glasses equipped with an egocentric six-channel microphone array, worn by one of the conversation participants. Figure \ref{fig:glasses} shows a drawing of the glasses with locations of the microphones \cite{donley2021easycom}.
\par 
The dataset contains recordings of natural conversations in a noisy restaurant environment. Participants were equipped with close-talk microphones, a camera and tracking markers. They were asked to engage in conversations during several tasks, including introductions, ordering food, solving puzzles, playing games, and reading sentences. The recordings also contain an egocentric video viewpoint of the participants. The pose (position and rotation) of every participant was also recorded. The dataset was additionally labelled with annotators of voice activity.
\par 
Altogether the dataset contains $nID\sim 346$ segments of approximately 1-minute duration. Each segment is given an index $ID$, $ID\in\{1,2,\ldots,nID\}$.
The signals recorded by the microphones were sampled at a rate of $48 \textrm{ kHz}$ which were then down-sampled to $16 \textrm{ kHz}$ . The recorded data was transformed into the STFT domain using a 1024 samples ($\simeq 64 \textrm{ msec}$) Hann window with an overlap of 512 samples. The microphone signals in the STFT domain were employed as an input to the algorithms under study.

Three  characteristic parameters of the  EasyComm dataset and the microphone array that are important in the context of this work, are:
\begin{itemize}
    \item
{\bf Time interval} $\Delta T$. This is the time over which the scenario may be regarded as spatially static and is investigated in Sect. \ref{SECT_time_interval}.
    \item 
{\bf Effective Frequency Range} $[f_{\text{low}},f_{\text{high}}]$. This is the effective frequency band over which the array presents high directivity but is also aliasing-free, and is analyzed in Sect. \ref{SECT_freq_range}. 

\end{itemize}
 
Before dynamic DOA experiments can be performed, it is necessary to optimally choose these parameters. This is detailed in the following sections. 

\subsection{\texorpdfstring{Selection of Time Interval $\Delta T$}{Selection of Time Interval Delta T}} \label{SECT_time_interval}

The EasyCom dataset consists of $nID\sim 346$ segments of  approximately 1-minute duration.  Each segment 
is divided into intervals of length $\Delta T$ seconds. We classify each interval as \lq\lq dynamic" if during this interval the change in ground truth position exceeds $\zeta$ degrees. Otherwise, the interval is \lq\lq static". Table \ref{tbl_zeeta} shows the percentage of \lq\lq dynamic" intervals for different angles $\zeta$ and different interval durations $\Delta T$. We see that the number of dynamic intervals is extremely low for all intervals $\Delta T \leq 500$ msec. The number of dynamic intervals increase somewhat at $\Delta T=1000$ msec. although the percentage of dynamic intervals remains low.
According to  these considerations we chose   $\Delta T=500$ msec. 
 
\begin{table}[h]
\centering
\caption{Percentage of \lq\lq dynamic" intervals for different change angles $\zeta$ and interval durations $\Delta T$}
\label{tbl_zeeta}
\begin{tabular}{ccccc} 
\toprule
\multirow{2}{*}{$\Delta T$ (msec)} & \multicolumn{4}{c}{change angle $\zeta$ (deg) } \\
\cmidrule(lr){2-5}
& 3 & 5 & 7 & 10\\
\midrule
100 & 00.0 \% & 00.0 \% & 00.0 \% & 00.0 \% \\
300 & 00.6 \% & 00.1 \% & 00.0 \% & 00.0 \% \\
500 & 02.4 \% & 00.5 \% & 00.1 \% & 00.0 \% \\
1000 & 07.6 \% & 02.5 \% & 00.9 \% & 00.2 \% \\
5000 & 30.6 \% & 17.0 \% & 09.0 \% & 03.5 \% \\
\bottomrule
\end{tabular}
\end{table}
\subsection{ Selection of Operating Frequency Range 
} 
\label{SECT_freq_range}
The EasyCom dataset involves speech sound which naturally limits the frequency range of interest \cite{monson2014perceptual}. This frequency band may be further reduced in practice by aliasing effects which arise
from the microphone array. The operating frequency range is found from the array directivity $\Lambda_{h}(\phi_{l},f_{k})$ (Eq. (\ref{EQN_STEERING+VECT_SIM})). 

\begin{figure}[H]
    \centering
    \includegraphics[width=0.49\linewidth]{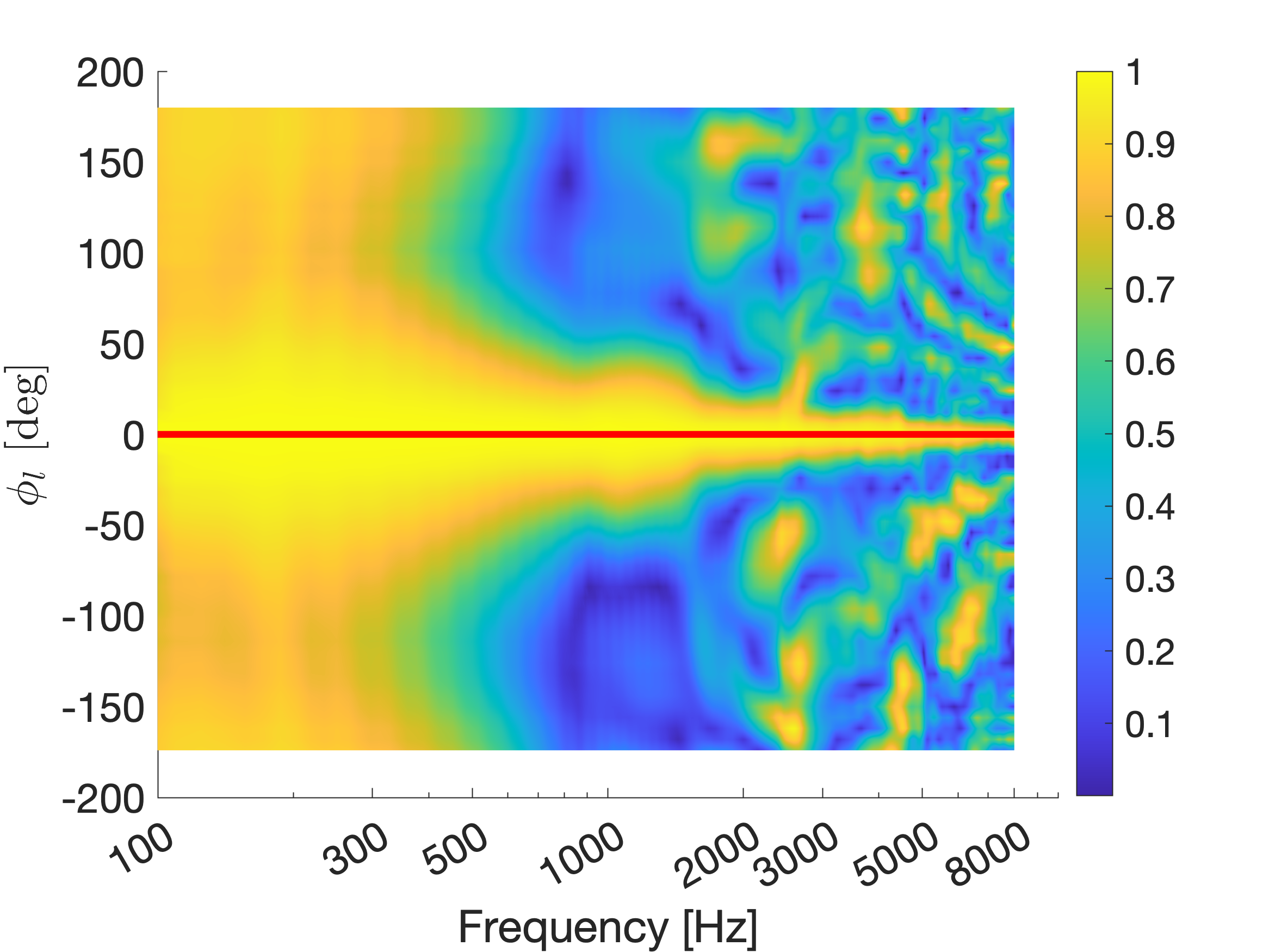}
    \caption{Array directivity as measured by Eq. (\ref{EQN_STEERING+VECT_SIM}), with $\phi_h=0^\circ$.}
    \label{fig:main2}
\end{figure}

Figure \ref{fig:main2} shows $\Lambda_{h}$ for look direction  $\phi_{h}=0^{\circ}$ and $M=4$ microphones. Visual inspection shows that  the preferred lower frequency is about $f_{\text{low}} = 1500$ Hz. For frequencies lower than this value, the directivity is too wide to be of any use. The situation for the upper frequency $f_{\text{high}}$ is less clear-cut.  For frequencies much greater than about $2500$ Hz, there are 
side lobes which \emph{may} degrade the spatial processing. This was investigated in a preliminary experiment. 
 The final choice of  $f_{\text{high}}$ was $3500$ Hz. 
 
\section{Preliminary Experiment for Parameter Selection}
 Several of the signal processing blocks in Fig. \ref{fig:flowchart-dynamic} have operating parameters which must be selected before DOA experiments can be performed. In this section, we show how these parameters were optimized. The parameters are:
\begin{itemize}
 
    \item The filter parameters used for smoothing the spectrum   ${\bf S}(t,f)$.
    \item The DPD threshold $\lambda$.
\end{itemize}
The experimental setup is given in section \ref{sub-ExperimentalSetupForParam} and the methodology used to optimize the parameters is given in Sect. \ref{SECT_methodology_para}. The experimental results for each parameter are given in Sects. \ref{sec:freqSmooth}-\ref{subsection_lambda}. 
\subsection{Experimental Setup} \label{sub-ExperimentalSetupForParam}
The operating parameters were experimentally optimized on the entire EasyComm dataset. This is described in Sect. \ref{sub-easycom}.

\subsection{Methodology}\label{SECT_methodology_para}
In optimizing the operating parameters we used two performance measures:
\begin{itemize}
    \item \textbf{Absolute DOA  Error}   $\varepsilon(t,f)$. 
This is defined for each time-frequency bin $(t,f)$ as the absolute difference between the true DOA $\Psi(t)$ and the estimated DOA $\hat{\theta}(t,f)$: 
\begin{equation}
\label{Defn_varepsilon}
\varepsilon(t,f)=|\Psi(t)-\hat{\theta}(t,f)|\;, 
\end{equation}
\par
\emph{Note}: in Eq.  (\ref{Defn_varepsilon}) we have assumed that both $\Psi(t)$ and $\hat{\theta}(t,f)$ are measured with respect to the same coordinate system. In practice,   $\Psi(t)$ is measured with respect to an  axes defined relative to the room, while $\hat{\theta}(t,f)$ is measured with respect to the orientation of the eyeglasses. Thus, before calculating $\varepsilon(t,f)$, we transform $\hat{\theta}(t,f)$  to the fixed axes of the room by incorporating head tracking information.

\item \textbf{Fraction of  Low Error Bins}
$n_{\text{low}}(T)$.  This is defined for each active interval $T$ as the fraction of valid bins with
low errors:
\begin{equation}
n_{\text{low}}(T)=\frac{1}{N_\text{val}(T)}\sum_{(t,f)_{T}}L(t,f)V(t,f)\;,
\end{equation}
where
\begin{align}    
L(t,f)&= 
\begin{cases}
    1 & \text{if } \varepsilon(t,f)\le 10^{\circ}\;,\\
    0              & \text{otherwise}\;,
\end{cases}\\ 
N_\text{val}(T)&=\sum_{(t,f)_{T}}V(t,f)\;,
    \label{eq_NN}
\end{align}
\end{itemize}

\subsection{Time and Frequency Smoothing}
 \label{sec:freqSmooth}
Reference \cite{beit2020importance} have shown that it is beneficial to smooth the directional spectrum
$\textbf{S}(t,f)$ before calculating the DOA estimates $\hat{\theta}(t,f)$ and the DPD values $\xi(t,f)$.
\par
For each bin $(t,f)$ we smooth $\textbf{S}(t,f)$  by averaging $\textbf{S}(t,f)$ in the range $t\in R_{t}$ and $f\in R_{f}$ around $(t,f)$.   
$\overline{\textbf{S}}(t,f)=[\overline{S}_{1}(t,f),\overline{S}_{2}(t,f),\ldots,\overline{S}_{L}(t,f)]^{T}$ denotes the smoothed spectrum, with
\begin{equation}
\overline{S}_{l}(t,f)=\frac{1}{ R_{t}R_{f}}\sum_{m=-r_{t}}^{r_{t}}\sum_{n=-r_{f}}^{r_{f}}S_l(t+m\delta T,f+n\delta f)\;, 
\label{eqSmootherSpectrum}
\end{equation}
 where $(\delta t,\delta f)$ is the size of STFT  bin,
$r_{t}=\lfloor (R_{t}/2)\rfloor$ and
$r_{f}=\lfloor (R_{f}/2)\rfloor$. 
 
\par 
Altogether we investigated four filters (listed in Table \ref{tblsmoothingFilters}) corresponding to different combinations of $(R_{t},R_{f})$ 
\begin{table}[h]
\centering
\caption{Smoothing Filters for $\textbf{S}(t,f)$}
\label{tblsmoothingFilters}
\begin{tabular}{ccc} 
\toprule
Filter Name &$R_{t}$ & $R_{f}$  \\
\midrule
U11 &1 & 1 \\
U33&3 & 3  \\
U37&3 & 7  \\
U57&5 & 7   \\
\bottomrule
\end{tabular}
\end{table}
where $U11$ corresponds to no-smoothing.
\par
\begin{figure}[H]
    \centering
  \includegraphics[width=0.49\linewidth]{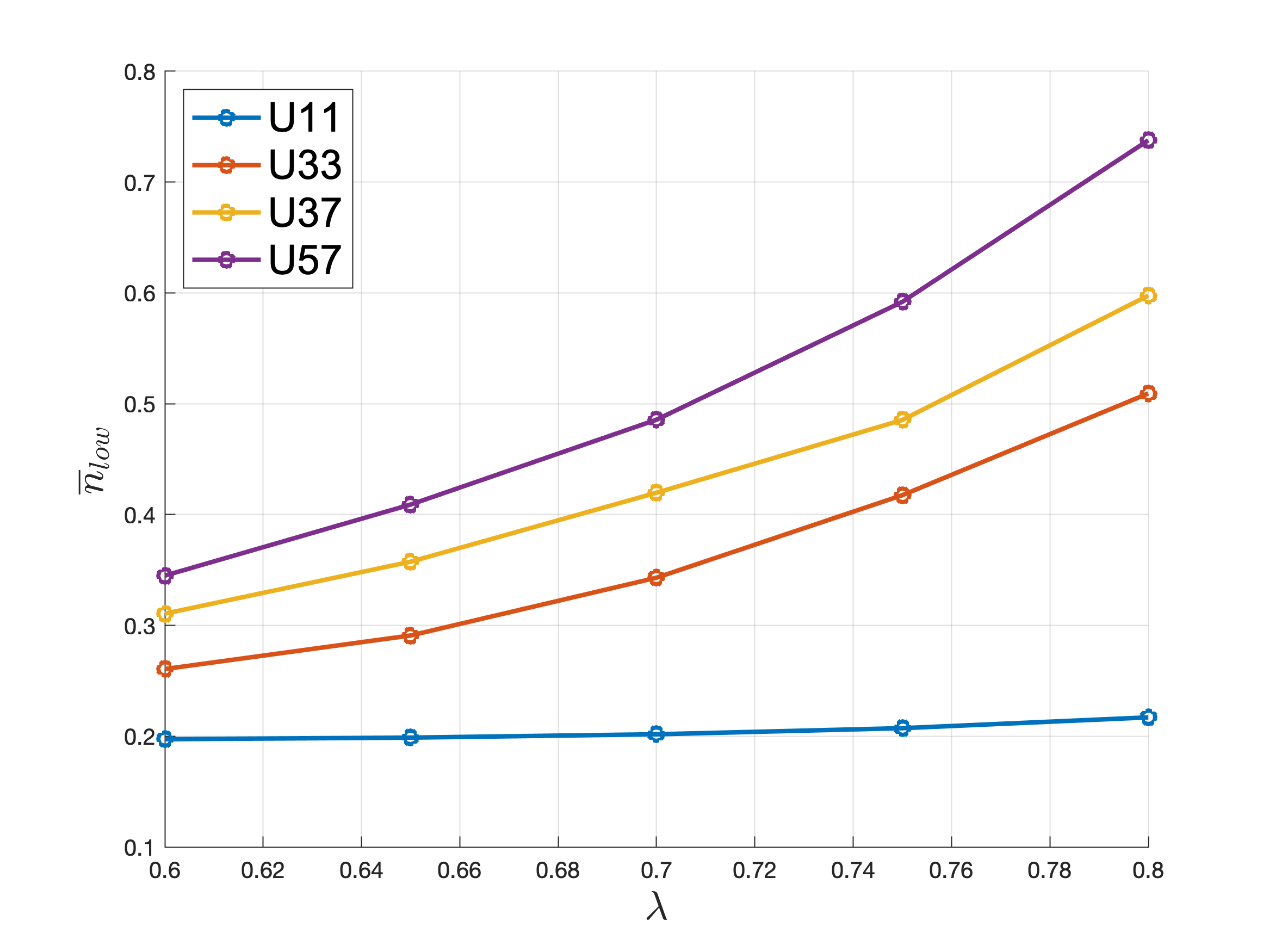} 
 \caption{Shows $\overline{n}_{\text{low}}$ as a function of $\lambda$ and  
   for   smoothing filters $U11$, $U33$, $U37$, $U57$. }  
   \label{fig:nlow-vs-lambda}
\end{figure}

Let $\overline{n}_{\text{low}}$ denote $n_{\text{low}}(T)$ averaged over all data sessions and all time periods $T$. Then
Fig. \ref{fig:nlow-vs-lambda} 
shows how $\overline{n}_{\text{low}}$ 
varies with  $\lambda$ for different  smoothing filters.    We see that $\overline{n}_{\text{low}}$ 
increases significantly with $\lambda$. 
Inspection of $\overline{n}_{\text{low}}$ shows large differences between the different smoothing filters for large $\lambda$. However,  these differences tend to decrease as $\lambda$ decreases. The final filter chosen was $U37$.

\subsection{\texorpdfstring{DPD Threshold $\lambda$}{DPD Threshold lambda}} \label{subsection_lambda}

\begin{figure}[H]
    \centering
  \includegraphics[width=0.49\linewidth]{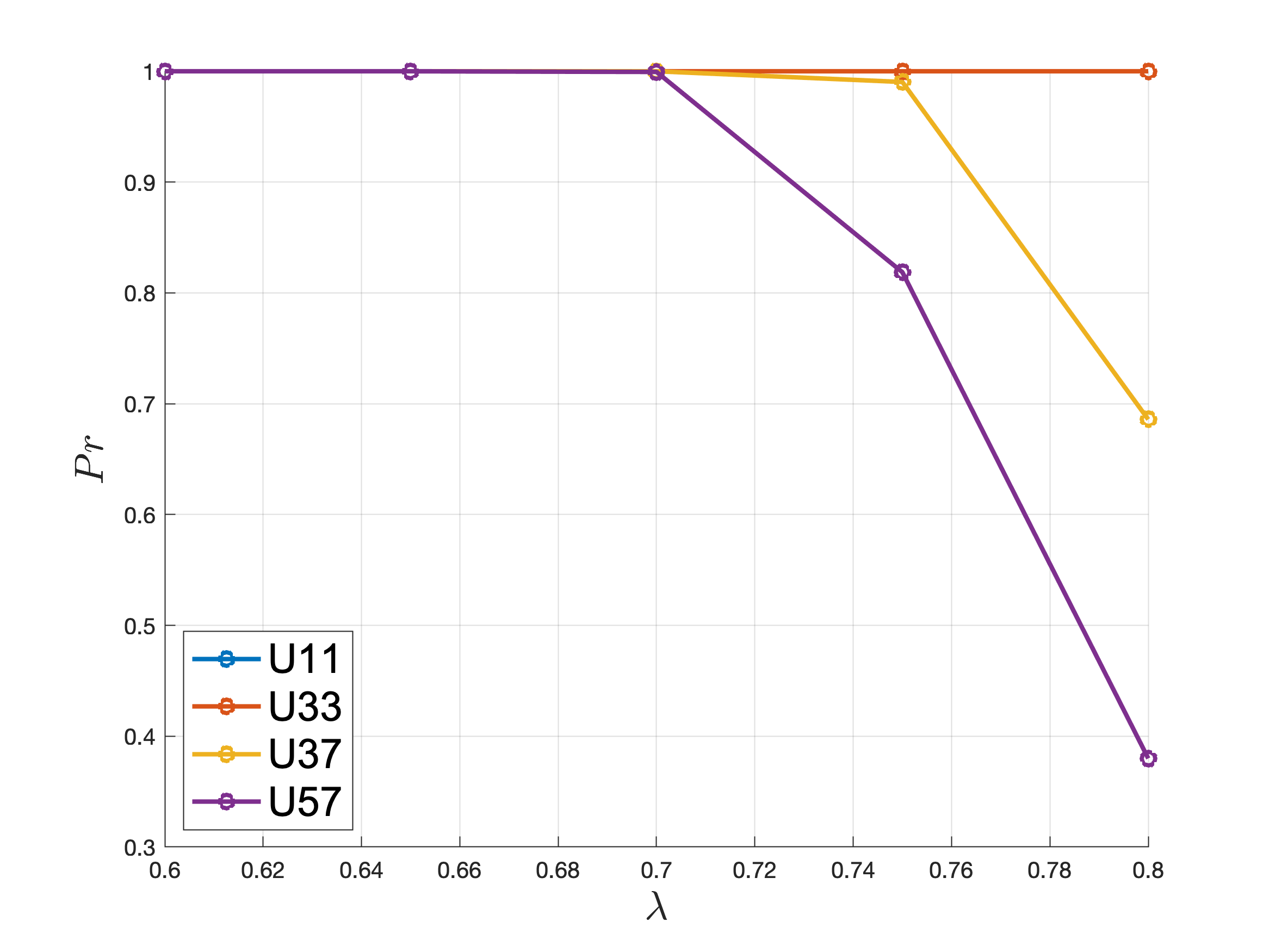} 
       \caption{Shows $Pr$ as a function of $\lambda$ for $\Delta T= 480$ msec and $f_{\text{high}}=3500$ Hz. The curves were calculated for different smoothing filters.
       }
    \label{fig:miss-vs-lambda}
\end{figure}

In optimizing $\lambda$, our two  principal requirements are:
\begin{itemize}
\item The probability $Pr$ is high, where $Pr$ is defined as the fraction of active intervals  which contain at least one valid time-frequency bin (see Eq. (\ref{eq:identity})) to the total number of active intervals. Active intervals are defined as intervals for which one or more speakers are speaking throughout the interval. 
    \item For most active intervals $T$ the proportion of valid bins corresponding to direct sound leading to accurate DOA estimates is high.. We may use the value of $\overline{n}_{\text{low}}$ as an estimate of this proportion. The corresponding curves are shown in Fig. \ref{fig:nlow-vs-lambda}
\end{itemize}
Fig \ref{fig:miss-vs-lambda}  
shows how $Pr$ 
varies with $\lambda$ for  several smoothing filters. We see that  $Pr$ decreases rapidly with $\lambda$. Inspection of these curves show a suitable value for $\lambda$ is $\lambda=0.7$. 

\par
  
\section{DOA Estimation Experiments}\label{SECT_results}
In this section, DOA estimation results obtained with the EasyCom dataset using the new DOA estimation algorithm LSDD-wQ are presented.  In particular, the focus is on measuring the differences in DOA performance between new and baseline algorithms as a result of using new weights $w_{\text{new}}(t,f)$ and a new quality measure $Q_{\text{new}}(k)$.

The experimental setup is described in Sect. \ref{sub-ExperimentalSetupFinal}. The methodology used is detailed in Sect. \ref{sect_methodology_final}. Subsequently, in section \ref{sect_results}, the experimental results obtained with the new DOA estimation algorithm are presented and  compared with those obtained using the baseline algorithm. 

\subsection{Experimental Setup} \label{sub-ExperimentalSetupFinal}
Both the new and the baseline DOA algorithms
were evaluated  on the entire EasyCom dataset (Sect. \ref{sub-easycom}). 

\subsection{Methodology}\label{sect_methodology_final}
The performance of the DOA estimation algorithms in this dynamic environment, and specifically, their accuracy and robustness,  are measured through a statistical analysis of   the  DOA estimates $\hat{\Theta}_{T}(k)$, their quality measures $Q_{T}(k)$ and  their errors $E_{T}(k)$ for $k\in\{1,2,\ldots,K\}$. Here, $\hat{\Theta}_{T}(k)$, $Q_{T}(k)$ and $E_{T}(k)$ represent the final DOA estimate, the quality measure, and the absolute error of the $k$th active speaker in the interval $T$, respectively. With $\Psi_{k}$ denoting the true DOA of the $k$th speaker in interval $T$,  the absolute error is given by:
\begin{equation}
    E_{T}(k)=|\widehat{\Theta}(k)-\Psi_{k}|\;.
\end{equation}

We compare the accuracy and robustness of the new and baseline DOA estimation algorithms  as follows:
for each algorithm we separately create a list of the absolute errors $E_{T}(k)$ and qualities $Q_{T}(k)$ for all $T$ and $k$. From each list we select a subset of $P\%$ errors, which have the highest $Q_{T}(k)$ values. Then   the mean absolute error of each subset is a measure of the estimation accuracy of the algorithm.  Similarly, the relative number of outliers, which we define as  absolute errors greater than $25^{\circ}$, is a measure of the robustness of the DOA algorithm. 
\par 
In a subset containing $P\%$ errors, the  mean absolute error  obtained with the new algorithm is denoted
as 
$\widebar{M}_{P}(w_{\text{new}},Q_{\text{new}})$. The corresponding mean absolute error obtained with the baseline algorithm is denoted as $\widebar{M}_{P}(w_{\text{base}},Q_{\text{base}})$. Likewise, $n_{P}(w_{\text{new}},Q_{\text{new}})$ and $n_{P}(w_{\text{base}},Q_{\text{base}})$ denote, respectively, the  fraction of outliers in a subset containing $P\%$ errors  obtained with the new and baseline algorithms. 
\par 
For comparison purposes, the \lq\lq ideal" mean absolute error for each $P$ value is also presented. This ideal mean absolute error is calculated assuming an oracle which selects the subset with the \emph{smallest} errors for each value of $P$. The  corresponding \lq\lq ideal" mean absolute errors obtained with the new and baseline algorithms are denoted as 
 $\widebar{M}_{P}(w_{\text{new}},Q_{\text{ideal}})$ and $\widebar{M}_{P}(w_{\text{base}},Q_{\text{ideal}})$, respectively.
\par

\subsection{Results}\label{sect_results}
In this section we compare  the experimental results obtained with the new and baseline algorithms. We divide the section into three: starting with the relative accuracies of the two algorithms and then discussing the robustness of the  algorithms. Finally, we discuss the relative contributions of $w_{\text{new}}$ and $Q_{\text{new}}$.
\par
\begin{figure}[H] 
\centering
    \includegraphics[width=0.75\linewidth]{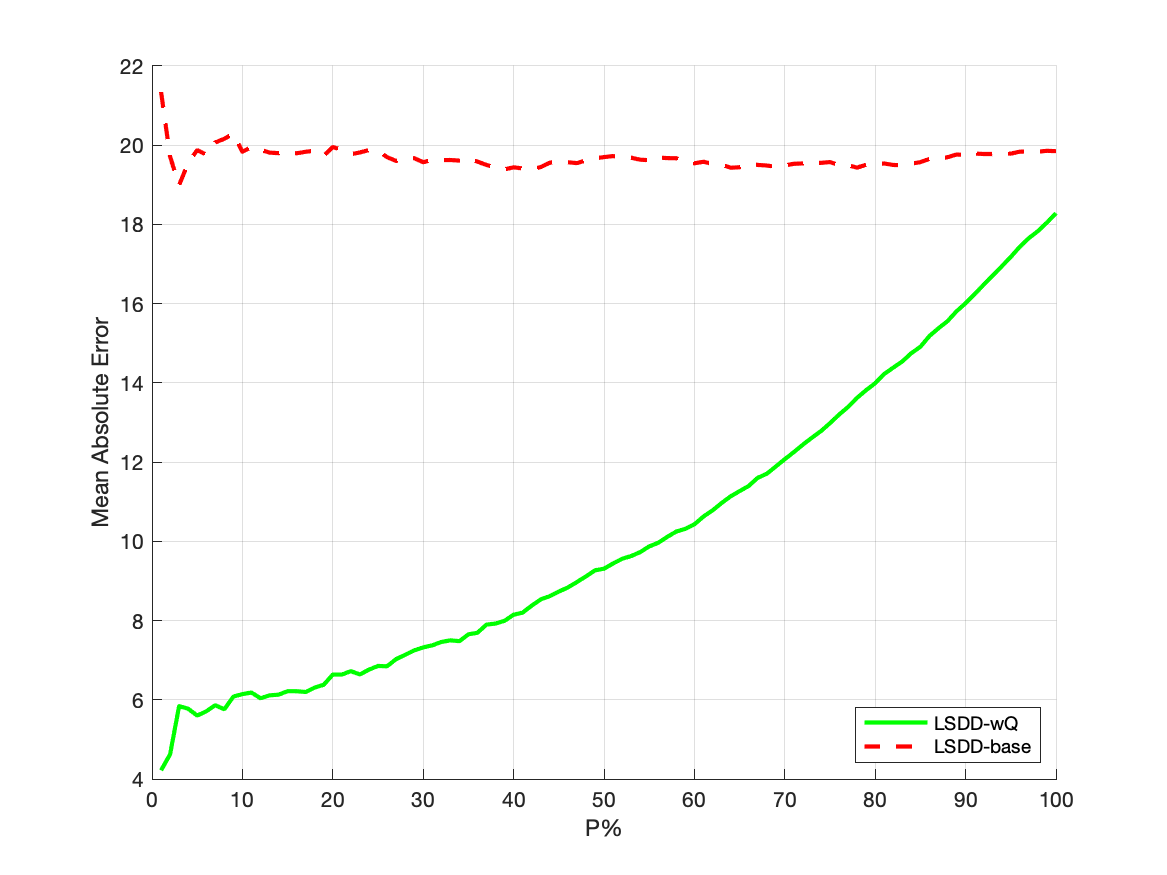}
       \caption{Shows the mean absolute errors $\widebar{M}_{P}(w_{\text{new}},Q_{\text{new}})$ and 
       $\widebar{M}_{P}(w_{\text{base}},Q_{\text{base}})$ as a function of $P\%$.   
       } 
 \label{fig:newvsbase_error}       
\end{figure}

\bigskip\noindent
{\bf Accuracy}\\
Fig. \ref{fig:newvsbase_error} shows how the mean absolute errors  $\widebar{M}_{P}(w_{\text{new}},Q_{\text{new}})$ and $\widebar{M}_{P}(w_{\text{base}},Q_{\text{base}})$ vary with $P$. 
We  observe how,  for all $P$, including $P=100\%$,  $\widebar{M}_{P}(w_{\text{new}},Q_{\text{new}})$  
is \emph{consistently} lower than 
$\widebar{M}_{P}(w_{\text{base}},Q_{\text{base}})$. 
Furthermore, we see that while the $\widebar{M}_{P}(w_{\text{base}},Q_{\text{base}})$ curve is relatively flat, the $\widebar{M}_{P}(w_{\text{new}},Q_{\text{new}})$ curve falls as $P$ decreases. For example, at $P=50\%$ the mean absolute error of LSDD-wQ drops to under $10^{\circ}$, while the corresponding baseline error   
 remains relatively constant at about $20^{\circ}$. 
 \par We thus conclude that for all values of $P$ the new algorithm LSDD-wQ is more accurate than LSDD-base with the relative accuracy of LSDD-wQ sharply increasing as we choose smaller values of $P$. 
 \par
\begin{figure}[H] 
\centering
    \includegraphics[width=0.75\linewidth]{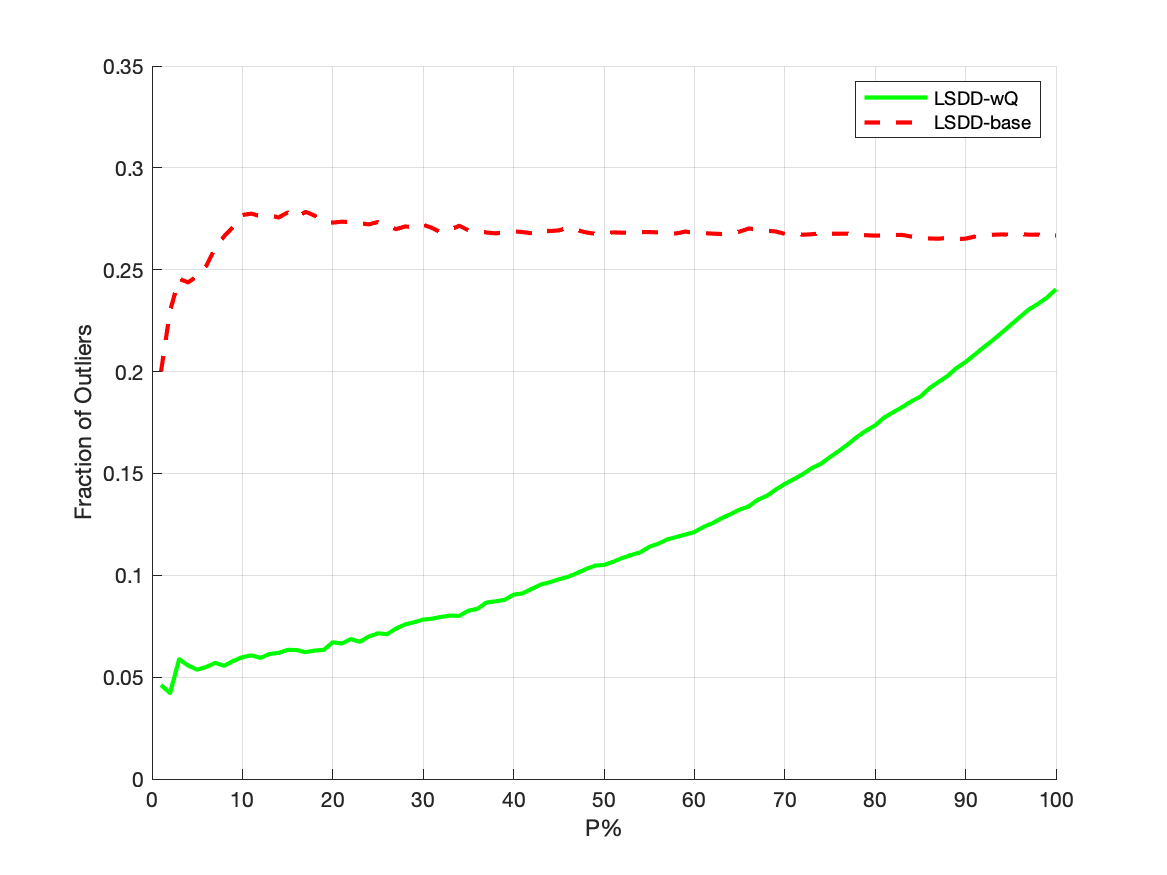}
       \caption{Shows the  fraction of outliers $n_{P}(w_{\text{new}},Q_{\text{new}})$ and
       $n_{P}(w_{\text{base}},Q_{\text{base}})$ as a function of $P\%$.       
       } 
 \label{fig:newvsbase_outlier}      
\end{figure}

\bigskip \noindent
{\bf Robustness}\\
 Fig \ref{fig:newvsbase_outlier}, shows how the fraction of outliers changes with $P$. We see clearly how for LSDD-wQ the fraction of outliers decreases sharply  as  $P$ is reduced. For instance, at $P=50\%$ the fraction of  outliers drops to approximately $10\%$, while the fraction of outliers in the baseline algorithm remains relatively constant at around $27\%$. 
\par
 We  thus conclude that the new algorithm LSDD-wQ is  substantially more robust than baseline algorithm LSDD-base. 
 
\par
\begin{figure}[H]
    \centering
    \includegraphics[width=0.75\linewidth]{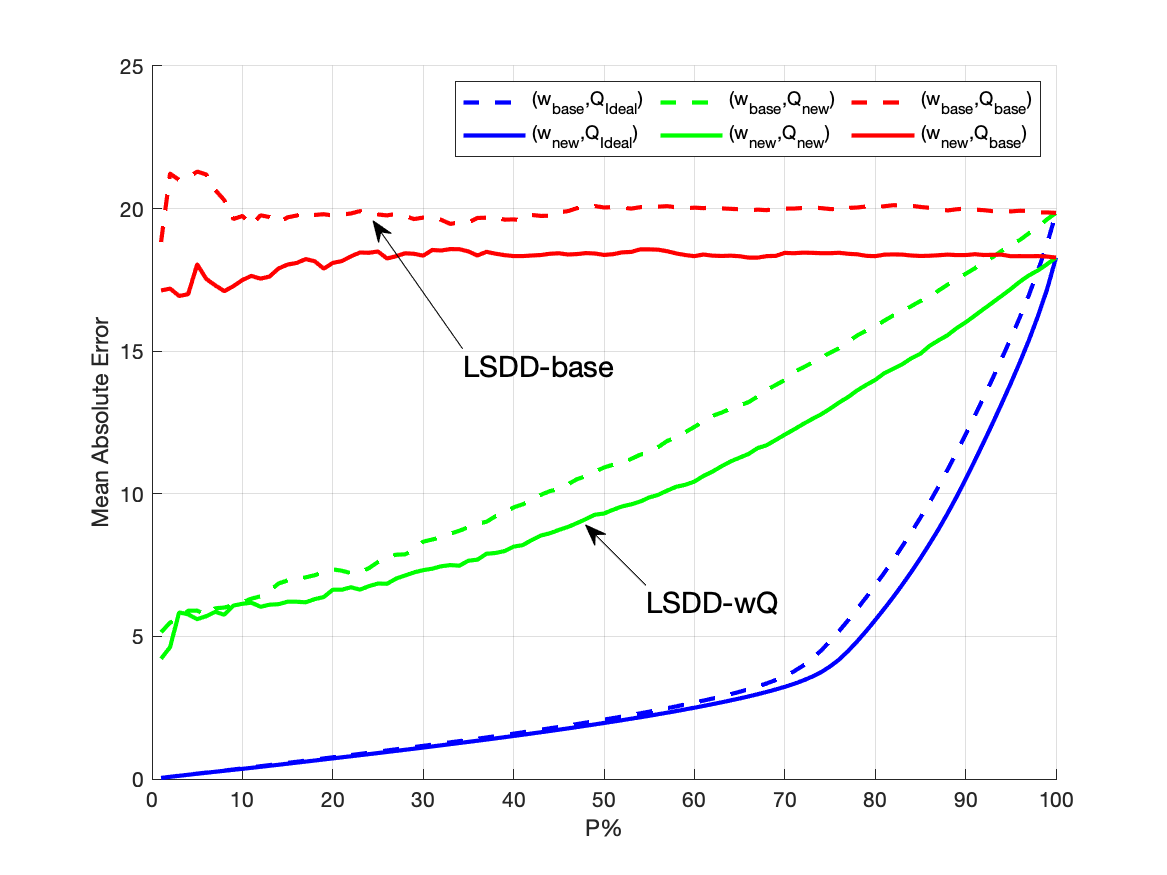}
       \caption{Shows the mean absolute errors $\widebar{M}_{P}(w_{\text{base}},Q_{\text{base}})$, 
       $\widebar{M}_{P}(w_{\text{new}},Q_{\text{base}})$, \\  
       $\widebar{M}_{P}(w_{\text{base}},Q_{\text{new}})$, 
       $\widebar{M}_{P}(w_{\text{new}},Q_{\text{new}})$, 
       $\widebar{M}_{P}(w_{\text{base}},Q_{\text{ideal}})$ and  
       $\widebar{M}_{P}(w_{\text{new}},Q_{\text{ideal}})$ \\
       as a function of $P\%$.}
    \label{fig:errorvspercentiles_ALL}
\end{figure}

\bigskip\noindent
{\bf Relative Contributions of $w_{\text{new}}$ and $Q_{\text{new}}$}\\
We now consider  the relative contribution of the weights $w_{\text{new}}$ and the quality measures $Q_{\text{new}}$ in the new algorithm LSDD-wQ. For this purpose, we introduce two \lq\lq combination" DOA algorithms:

\begin{itemize}
\item 
The first combination algorithm uses 
new weights $w_{\text{new}}(t,f)$  with baseline  cluster quality  $Q_{\text{base}}(k)=1$. In  this case, for each cluster $k$, we have a DOA estimate $\hat{\Theta}(k)$ and a quality measure $Q(k)$, where 
\begin{equation}\label{family_LSDD_1_combA}
\begin{aligned}
\begin{cases}
    \widehat{\Theta}(k)&=\widehat{\Theta}_{\text{new}}(k)  \\
    Q(k)&=Q_{\text{base}}(k) \;,
    \end{cases}
    \end{aligned}
    \end{equation}
    The mean absolute error of a subset $P\%$ obtained using this algorithm is $\widebar{M}_{P}(w_{\text{new}},Q_{\text{base}})$ and the corresponding fraction of outliers is $n_{P}(w_{\text{new}},Q_{\text{base}})$.
\item 
 The second combination algorithm uses baseline weights $w_{\text{base}}(t,f)=1$ with new cluster quality measures $Q_{\text{new}}(k)$. In this case, for each cluster $k$,  
 we have a DOA estimate $\hat{\Theta}(k)$ and a quality measure $Q(k)$, where 
\begin{equation}\label{family_LSDD_2_combA}
\begin{aligned}
\begin{cases}
    \widehat{\Theta}(k)&=\widehat{\Theta}_{\text{base}}(k)  \\
    Q(k)&=Q_{\text{new}}(k) \;,
\end{cases}
\end{aligned}
\end{equation}
    The mean absolute error of a subset $P\%$ obtained with this algorithm is $\widebar{M}_{P}(w_{\text{base}},Q_{\text{new}})$. The corresponding fraction of outliers is $n_{P}(w_{\text{base}},Q_{\text{new}})$. 
\end{itemize}
  Fig. \ref{fig:errorvspercentiles_ALL} shows how the mean absolute values 
 obtained with the new, the baseline, and the two combination algorithms vary with $P$. In this figure we also show  the \lq\lq ideal" curves
 $\widebar{M}_{P}(w_{\text{new}},Q_{\text{ideal}})$ and
 $\widebar{M}_{P}(w_{\text{base}},Q_{\text{ideal}})$. 
\par
It is first observed that there is a natural ordering of the curves. For convenience we divide  the curves into three pairs which we identify  by coloring them  red, green and blue.  The red, green and blue curves, use, respectively  baseline quality $Q_{\text{base}}$, new quality $Q_{\text{new}}$ and ideal quality $Q_{\text{ideal}}$ measures. However, each curve in the pair  uses different weights. We  use  dashed lines to represent curves using  baseline weights $w_{\text{base}}$ and  full lines to represent curves using the new weights $w_{\text{new}}$. 
\par
We see clearly that the baseline quality  (red) curves
lie above the new quality  (green)  curves 
which in turn lie  above the ideal quality (blue) curves. Moreover, the difference in mean absolute value between the green and blue curves falls to about $5^{\circ}$ as we decrease $P$ while the difference in mean absolute value between the green and red curves reaches as much as $ 15^{\circ}$.  
\par
On the other hand, we clearly see for each pair of curves using the same quality measure, that the difference in mean absolute value is around $2-3^{\circ}$ or less. 
\par
We thus conclude that the new quality measure $Q_{\text{new}}$ makes the largest contribution in improving the DOA accuracy of the LSDD-wQ algorithm. Moreover, as we reduce $P$  the difference between curves representing  the new quality measure and ideal quality measure fall to about $5^{\circ}$. 

\par

\section{Conclusion}
A detailed investigation into DOA estimation, particularly through algorithms based on LSDD, has been conducted. A novel methodology was described for performing DOA estimation in dynamic scenarios with wearble arrays. This included selection of an effective operating frequency range,  developing a novel subtractive weighted clustering algorithm with quality measure for each cluster and giving each time-frequency estimate a reliability weight. The introduction of the new reliability weight gave a small, but consistent, improvement in DOA accuracy. The introduction of a  cluster quality measure  was found to be highly correlated with the DOA estimation accuracy. In particular, the cluster quality can be used to reduce outliers \emph{before} tracking is undertaken. This in turn can make tracking algorithms both more accurate and robust.

\section{Acknowledgement}
This work was partially supported by Reality Labs Research @ meta

\bibliography{sn-bibliography}

\end{document}